\documentclass[sigconf, nonacm]{acmart}

\AtBeginDocument{%
  \providecommand\BibTeX{{%
    \normalfont B\kern-0.5em{\scshape i\kern-0.25em b}\kern-0.8em\TeX}}}

\usepackage{float}
\usepackage{enumitem}
\usepackage{listings}

\begin{document}

\title{The Data Lakehouse: Data Warehousing and More}

\author{Dipankar Mazumdar}
\email{dipankar.mazumdar@dremio.com}
\orcid{1234-5678-9012}
\affiliation{%
  \position{Developer Advocate}
  \institution{Dremio Inc}
  \streetaddress{50 Power St}
  \city{Toronto}
  \state{Ontario}
  \country{Canada}
  \postcode{M5A 0V3}
}

\author{Jason Hughes}
\email{jason@dremio.com}
\affiliation{%
  \position{Director, Technical Advocacy}
  \institution{Dremio Inc}
  \streetaddress{P.O. Box 1212}
  \city{San Diego}
  \state{California}
  \country{USA}
}

\author{JB Onofré}
\email{jb.onofre@dremio.com}
\affiliation{%
  \position{Principal Software Engineer}
  \institution{Dremio Inc}
  \city{Henvic}
  \state{Finistère}
  \country{France}
}


\begin{abstract}
Relational Database Management Systems designed for Online Analytical Processing (RDBMS-OLAP) have been foundational to democratizing data and enabling analytical use cases such as business intelligence and reporting for many years. However, RDBMS-OLAP systems present some well-known challenges. They are primarily optimized only for relational workloads, lead to proliferation of data copies which can become unmanageable, and since the data is stored in proprietary formats, it can lead to vendor lock-in, restricting access to engines, tools, and capabilities beyond what the vendor offers.


As the demand for data-driven decision making surges, the need for a more robust data architecture to address these challenges becomes ever more critical. Cloud data lakes have addressed some of the shortcomings of RDBMS-OLAP systems, but they present their own set of challenges. More recently, organizations have often followed a two-tier architectural approach to take advantage of both these platforms, leveraging both cloud data lakes and RDBMS-OLAP systems. However, this approach brings additional challenges, complexities, and overhead.

This paper discusses how a data lakehouse, a new architectural approach, achieves the same benefits of an RDBMS-OLAP and cloud data lake combined, while also providing additional advantages. We take today's data warehousing and break it down into implementation-independent components, capabilities, and practices. We then take these aspects and show how a lakehouse architecture satisfies them. Then, we go a step further and discuss what additional capabilities and benefits a lakehouse architecture provides over an RDBMS-OLAP.

\end{abstract}

\ccsdesc[500]{Information systems~Database management system engines}
\ccsdesc[500]{Information systems~Database design and models}

\keywords{database, OLAP, OLTP, data warehouse, data lakes, data lakehouse, table format}


\maketitle

\section{Introduction}
Data architectures have evolved rapidly to support analytical workloads \cite{10037358}. With the increase in the number of applications generating data, enterprises needed a more efficient way to manage and make sense of all the data to ultimately help make business decisions. Additionally, the range of use cases has expanded, as enterprises now seek to run diverse workloads, including both dashboards and machine learning (ML) tasks, on the same copy of the data.

Relational databases optimized for Online Transactional Processing (RDBMS-OLTP) \cite{10.1145/1376616.1376713} have long been a standard way for organizations to store and retrieve transactional data. They power OLTP-based applications to insert, update, or delete transactions in real time. For instance, booking a new flight ticket online is a type of transaction that would add rows to a database table containing booking details. OLTP databases are built for this usage pattern, which involves handling one or a few rows at once. However, if you want to analyze aggregated data, such as obtaining the total number of flight bookings over a period of time, using databases designed for OLTP leads to critical performance issues, even at a small scale. This led to the advent of database management systems (DBMS) designed and optimized for OLAP\cite{10.1145/248603.248616}. An RDBMS-OLAP is a data system used for storing and analyzing large amounts of data. RDBMS-OLAPs are preferred over RDBMS-OLTPs for OLAP workloads for a few reasons, including storing the data in a columnar format, the compute and storage engines taking advantage of the columnar layout, and usually having a Massively Parallel Processing (MPP) architecture\cite{INMON201557}. Data often comes in from a multitude of sources, e.g., application databases, CRMs, etc., in a structured format. It makes it easier for workloads such as business intelligence (BI) to refer to a single source of data for deriving insights. They are also more commonly known as Data Warehouses\cite{kimball2011data}.

The term “data warehouse” has been an overloaded one. Historically, it has been used to refer to two different things: the underlying technology itself and the technology-independent practices typically implemented in conjunction with the technology. To ensure reader clarity and clear distinction on these two aspects, we refer to the technology as RDBMS-OLAP and the technology-independent approaches as data warehousing practices (DW practices) in this paper. Furthermore, we refer to the combination of the technology used (RDBMS-OLAP) and the practices implemented in conjunction with the technology (DW practices) as Data Warehousing throughout this paper.

\begin{figure}
  \centering
  \includegraphics[width=\linewidth]{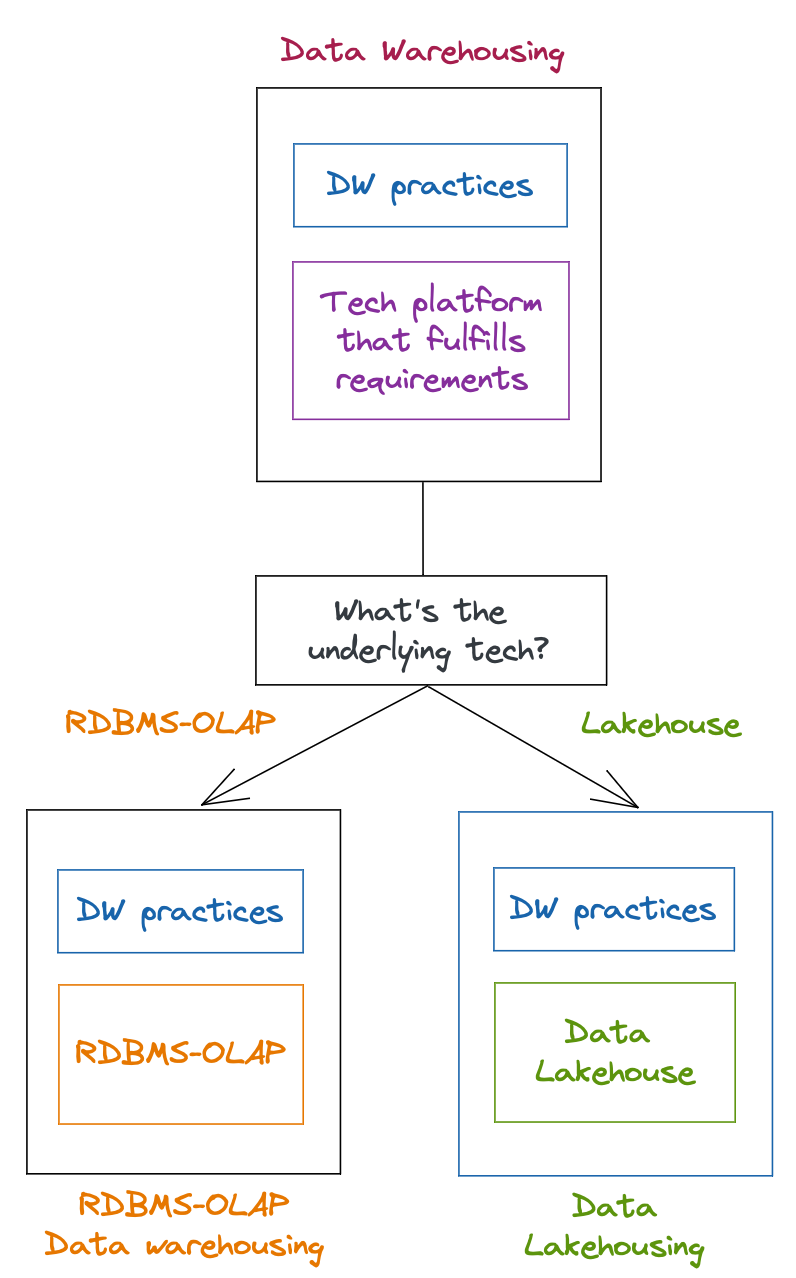}
  \caption{What is Data Warehousing?}
  \label{fig:figure_dwh}
\end{figure}
\setlength{\textfloatsep}{10pt}
In a more generic sense, if we have a technology capable of supporting an organization’s OLAP workload requirements and the technology-independent practices applicable to it, a data warehousing system can be agnostic of the underlying technology. For example, OLTP databases\cite{Insausti2019} have also been used as data warehouses at small scale because they provide the capabilities required and enable the use of DW practices. With this in mind, we can holistically define data warehousing based on the technology as shown in Figure ~\ref{fig:figure_dwh}.

Data Lakehousing is an emerging approach that brings all the capabilities of data warehousing to a data lake. A lakehouse\cite{Zaharia2021LakehouseAN} is not just a mere integration between the two data platforms — the design goal is to combine the advantages of the two technologies while eliminating their disadvantages. This paper argues that the platform capabilities and technology-independent practices that are part of the data warehousing approach can also be achieved in a data lakehouse. Moreover, data lakehousing addresses and enables certain aspects where data warehousing falls short, including advanced analytical workloads like machine learning. \\

\begin{tabular}{|p{0.3\linewidth}|p{0.3\linewidth}|p{0.3\linewidth}|}
  \hline
  \textbf{Tech-Independent Practices} & \textbf{Technology} & \textbf{Collective Term} \\
  \hline
  DW practices (e.g., data modeling, data quality, etc.) & RDBMS-OLAP & RDBMS-OLAP Data Warehousing \\
  \hline
  DW practices (e.g., data modeling, data quality, etc.) & Data Lakehouse & Data Lakehousing \\
  \hline
\end{tabular} \\

In the following sections, we define the requirements of RDBMS-OLAP data warehousing, introduce the data lakehouse architecture, describe how it addresses the requirements of data warehousing, and discuss the additional value it brings in terms of being an open data architecture. 
\section{Requirements of Data Warehousing}
In this section we discuss the various requirements that define RDBMS-OLAP data warehousing. The requirements are categorized into three subsections.
\begin{enumerate}
    \item Technical components
    \item Technical capabilities
    \item Technology-independent practices
\end{enumerate}

\subsection{Technical Components (RDBMS-OLAP Components)}
\label{section:tc}
\begin{figure}
  \centering
  \includegraphics[width=\linewidth]{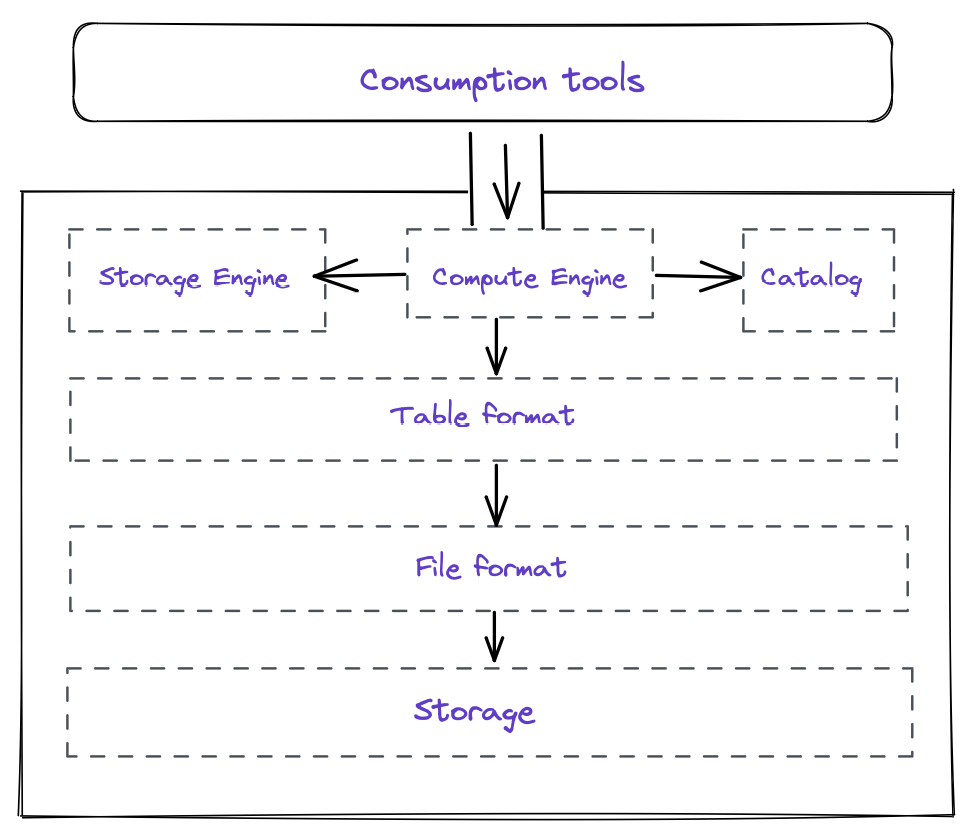}
  \caption{A generic representation of an RDBMS-OLAP data warehouse. Note that all the technical components are bundled into a single unit.}
  \label{fig:figure_dwh_arc}
\end{figure}

The technical components of an RDBMS-OLAP comprise the fundamental building blocks that underpin the infrastructure of a RDBMS-OLAP data warehouse. These components encompass the data storage system, file and table formats, compute and storage engines, etc. Understanding the technical components is crucial for building a reliable and efficient RDBMS-OLAP data warehouse architecture. Figure ~\ref{fig:figure_dwh_arc} presents an overview of these components.

\begin{enumerate}
\item \textbf{Data storage:} Data warehouses need to efficiently store vast volumes of data from different sources. The data storage\cite{10.5555/1083592.1083658} component of an RDBMS-OLAP plays a pivotal role in efficiently housing and organizing vast volumes of data from various sources. It is a way to take the bytes and store them at a specific path in the file system. By employing optimized storage mechanisms and structured file systems, it enables quick data access, retrieval, and manipulation for effective data analysis and reporting.

\item \textbf{File format:} The data storage component stores raw data as files within the file system. And, file formats determine how data is written within each file. Formats can be structured, semi-structured, or loosely structured, depending on the data type. They can store data in a row or columnar manner\cite{10.1145/1376616.1376712}, with columnar formats having performance advantages when dealing with large numbers of records at a time.

\item \textbf{Table format:} A table format serves as a metadata layer above the file format component, providing an abstraction that separates the logical representation from the physical files. They help an RDBMS-OLAP organize and manage the data and metadata files stored in the storage layer, allowing you to leverage techniques such as partitioning and clustering for efficient query processing. In simpler terms, a table format organizes a group of data files in the form of a table so it can be used for logically querying the data using the compute engine. 

\item \textbf{Storage engine:} A storage engine\cite{8820952} is responsible for organizing the data and keeping all the files and data structures up to date (e.g., indexes, non-null constraints, foreign key constraints). It facilitates user interactions and supports essential database operations like CRUD (create, read, update, and delete) through its APIs. The storage engine is designed to efficiently handle large volumes of structured and semi-structured data, optimizing performance and scalability. It implements various storage mechanisms and techniques to ensure fast data retrieval, data compression, and efficient data access.

\item \textbf{Compute engine:} A compute engine\cite{10.1145/2465351.2465355} executes queries to process the data stored in the storage layer. Along with query execution, it also handles data transformations, aggregations, and other computational tasks needed to generate insights from the stored data. The compute engine leverages various techniques for query acceleration and optimization to maximize performance and minimize query execution time. Modern cloud data warehousing systems\cite{10.1145/2882903.2903741} use MPP systems that enable highly concurrent workloads. MPP has been successfully deployed by systems such as Teradata\cite{Teradata} and Netezza\cite{Netezza}. Unlike SMP (Symmetric Multi Processing)\cite{SMP} which runs the workload in a single machine, MPP architecture deploys a cluster of machines, with compute distributed on the cluster. This kind of architecture is able to handle massive data volumes, but also be scalable on the computing layer as it’s possible to add additional nodes for processing. Modern cloud data warehouses like Snowflake are based on an MPP architecture.

\item \textbf{Catalog:} an RDBMS-OLAP stores datasets from a variety of sources. Users need to quickly identify the data they need for analysis. A catalog addresses these problems by using metadata\cite{doi:10.1142/S0218843001000357} such as table name, schema, etc., to identify datasets. This is sometimes also known as a data dictionary. Catalogs help in effective data management by categorizing data and making the process of data search, discovery, and accessibility easier. Note that native RDBMS-OLAP catalogs are distinct from external enterprise data catalogs and their capabilities. 

\end{enumerate}

\subsection{Technical Capabilities provided by RDBMS-OLAP}
\label{section:tcp}
RDBMS-OLAP implements a wide range of capabilities to enable efficient data management and analysis. These technical capabilities arise from one or more of the technical components above and collectively empower users to harness the full potential of a RDBMS-OLAP data warehouse, facilitating seamless data analysis, decision making, and knowledge discovery.

\begin{enumerate}
\item \textbf{Governance and security:} Since an RDBMS-OLAP is a central repository that stores data from various sources, it is imperative to have effective data governance strategies to ensure data is available, usable, and aligned to the security policies of an organization. Data governance streamlines accessibility of data so the right user group or tool can access the right data and that any sensitive information is protected. This can take the forms of row or column-level security, role-based access control, audit logging, etc. \cite{7804961} Additionally, RDBMS-OLAP implements various encryption techniques to prevent unauthorized access of data stored in the storage or during transfers. A lot of security regulations (i.e GDPR) are also relevant to geographic location, hence there needs to be capabilities in the RDBMS-OLAP to effectively deal with them.

\item \textbf{High concurrency:} Concurrent read and write operations on the same dataset is a very common scenario in data warehousing with the increasing analytical workloads. There might be multiple users accessing a dashboard that leverages the same data from the system at the same time while that dataset is being updated by another job. RDBMS-OLAP has the ability to scale out and provide the necessary performance and change isolation \cite{8509385} for multiple queries and users to execute tasks in parallel.

\item \textbf{Low query latency:} Query latency is a critical aspect for businesses running analytics such as BI and predictive analytics using data from an RDBMS-OLAP. Techniques such as query optimization\cite{PARK2002379}, indexing\cite{DBLP:conf/vldb/AgrawalCN00}, partitioning\cite{880634}, and caching mechanisms\cite{PARK2003103} are used by RDBMS-OLAP to facilitate faster queries for analytics and reporting.

\item \textbf{Ad hoc queries:} Executing ad hoc queries (single questions or multiple requests) are very common for analysts to answer a specific type of business question. For example, you may want to know why the sales of a product dropped in a specific month. RDBMS-OLAP provides support for these types of queries. 

\item \textbf{Workload management (WLM):} an RDBMS-OLAP system comprises various workloads such as loading data into the system, transforming data, and querying data. These workloads can be run concurrently by multiple users or tools. Hence it is necessary to have efficient workload management capabilities\cite{10.1145/3514221.3526045} to better manage the resources used for specific workloads and to avoid breaking any service-level agreements (SLAs) from a business perspective. WLM allows you to manage the cluster resources and workloads.

\item \textbf{Schema and physical layout evolution:} Business requirements are constantly changing and so is the data. Changes in the internal schema of a table to accommodate new requirements, e.g., adding a new column or altering an existing column in a table, is therefore a common scenario. RDBMS-OLAP enables evolving the schema\cite{10.1007/978-3-540-30464-7_48} with time and ensures there is no impact on anything else, assuming the changes made are backwards-compatible.

\item \textbf{ACID-compliant transactions:} Transactions play a crucial role in maintaining data integrity and consistency within an RDBMS-OLAP. They enable users to perform multiple operations as a single unit, ensuring atomicity and reliability. RDBMS-OLAP needs to guarantee that every read and write operation is reliable and consistent. Therefore, adhering to ACID (atomicity, consistency, isolation, durability) principles \cite{10.5555/2740975} is crucial.

\begin{enumerate}[label=$\circ$]
    \item \textbf{Atomicity}: statement (in a transaction) is a single unit, meaning that the entire statement is executed, or nothing.
    \item \textbf{Consistency}: ensures changes in transactions are performed as expected.
    \item \textbf{Isolation}: ensures that concurrent transactions (on the same table, for instance) don't interfere with each other.
    \item \textbf{Durability}: ensures that changes are persisted.
\end{enumerate}
\vspace{10pt}
ACID compliance ensures that each transaction is executed in its entirety and doesn’t introduce inconsistencies or erroneous data to downstream applications. By upholding ACID properties, RDBMS-OLAP maintains data integrity and ensures successful completion of transactions.

Key features in this context include:

\begin{itemize}
    \item Multi-statement \& Multi-table transactions: Users can execute a series of operations (such as inserts, updates, deletes) that modify multiple tables\cite{Big1} simultaneously. This allows for the cohesive management of related data changes, preventing data quality issues and ensuring accurate results. For instance, data load for a given point in time often isn’t for a single table, but rather a set of tables. If the changes for one table are visible downstream before the rest are, it can lead to inconsistencies. RDBMS-OLAP systems provide support for multi-statement transactions, facilitating comprehensive data quality checks.

    \item Transaction rollback: In the event of erroneous data or incorrect metrics, the ability to roll back a transaction becomes essential. For example, let’s say some newly ingested data leads to wrong metrics in a near-real-time operational dashboard. Since business users heavily depend on these dashboards to make a decision, it is imperative to have correct data. Transaction rollback is a fundamental feature of RDBMS-OLAP systems.

    \item Locking model: A robust locking mechanism\cite{10.1145/356842.356846} is vital for controlling concurrent data access and preventing conflicts or corruption. By managing read and write access to data, a locking model ensures data consistency. RDBMS-OLAP systems incorporate effective locking mechanisms to support concurrent transactions and maintain data integrity.
\end{itemize}

\item \textbf{Separation of storage and compute:} Traditional RDBMS-OLAP systems faced limitations in scalability and resource allocation due to the coupling of storage and compute components. The need for additional compute resources for certain queries necessitated scalable and uninterrupted resource scaling. However, this separation was not possible in traditional RDBMS-OLAP architectures. Cloud-based RDBMS-OLAP\cite{10.1145/2882903.2903741} has overcome this limitation by enabling independent scaling of storage and compute resources. This decoupling allows dynamic scaling of compute resources, ensuring optimal query performance.

\end{enumerate}

\subsection{Technology-Independent Practices (DW Practices)}
Technology-independent data warehousing practices encompass a range of methodologies that prioritize fundamental aspects such as data modeling\cite{10.1016/S0306-4379(02)00021-2}, Master Data Management (MDM)\cite{berson2011master}, referential integrity\cite{Watt11}, and Extract-Transform-Load (ETL) \cite{Vassili11} processes. These practices form the backbone of efficient data storage, organization, and use, regardless of the specific technology used. By adopting these practices, organizations can establish a robust foundation for their data architecture, enabling scalability, flexibility, and consistency in data management.

\begin{enumerate}
    \item \textbf{Data modeling:} Data modeling is the process of creating a model (logical or physical) for your organizational data that maps the relationship between various data objects stored in a particular storage. It helps standardize the data and translate user requirements into a technical form so it can be used to support and answer business-specific questions. Data modeling is logical as well as physical. A logical data model sets the foundation for a physical model by defining the data, entities, their attributes and relationships, etc. While the physical data model is the actual implementation of the data system. Data modeling techniques are agnostic of the data architectures (RDBMS-OLTP, RDBMS-OLAP).

    \item \textbf{ETL/ELT:} Extract-Transform-Load (ETL) is the process of extracting data from disparate data sources, transforming them, and loading them into a centralized repository such as an RDBMS-OLAP or data lake. Whether you want to load day-to-day operational data from your application databases or a CRM\cite{8117887}, ETL is foundational for most of the analytical use cases (i.e.,  BI and machine learning) in an organization. While ETL has traditionally been the most common way to move data between operational and analytical systems, another approach is Extract-Load-Transform (ELT) which involves extracting the data, loading it into a data system, and then processing it. Loading the data prior to transformation offers more flexibility in terms of storing various types of data (structured, semi-structured, etc.), leveraging a data system’s MPP capabilities, and processing only the data needed for a specific use case.

    \item \textbf{Data quality:} Data quality\cite{Cai-2015} is a fundamental aspect of effective data management. It ensures that data is accurate, consistent, and reliable for use in business processes and decision making. Three key practices, though not all, that contribute to data quality are:

    \begin{itemize}
        \item Master Data Management (MDM): MDM practices aim to maintain reliable and consistent information about critical organizational data assets, such as customer and product data. By centralizing and managing data from multiple sources, MDM mitigates issues like data fragmentation, duplication, and outdated copies, enabling businesses to rely on trustworthy data for their various use cases.

        \item Referential integrity: Referential integrity ensures the preservation of relationships between different data objects. This is crucial for data quality and usability, as changes in one object must be reflected in related objects to avoid data loss and inaccuracies. By enforcing referential integrity, organizations can maintain the consistency and reliability of their data.

        \item Slowly changing dimensions: In data modeling schemas like Kimball-Star, dimension tables play a significant role in analysis by providing descriptive attributes to fact tables. Slowly changing dimensions (SCDs)\cite{FAISAL2014151} address the challenge of managing changing attribute values over time. For example, a customer dimension table may contain attributes such as customer addresses and phone numbers, which can change over time, and SCDs help track and manage these changes while preserving historical data. Maintaining history is essential for analytical systems that rely on historical data for insights.
        
    \end{itemize}
\end{enumerate}

\section{Challenges with Data Warehousing}
\begin{enumerate}
    \item \textbf{Limited to structured data workloads:} RDBMS-OLAP was primarily designed to store data in a structured form (schema-on-write), so dashboards and analytical applications could leverage it. Now, as the analytics journey of an organization matures, the need to go beyond structured data is imperative. The availability of semi-structured (e.g., JSON) and loosely structured (e.g., images) data opens the scope for more advanced analytical workloads such as machine learning, natural language processing\cite{Zaharia2021LakehouseAN}, etc. However, since an RDBMS-OLAP is optimized for primarily structured data, businesses have to rely on another form of architecture called data lakes\cite{refId0} to store data that is not limited to the type. Moreover, machine learning models often demand iterative training on large datasets for effective results, which can pose challenges within a data warehousing environment due to limitations in scalability, processing model, processing power, and cost. Therefore, data scientists resort to accessing data directly from a data lake where data is stored in open file formats, such as Apache Parquet\cite{Vohra2016}, enabling them to perform exploratory analysis, feature engineering, and model training in a more agile way using engines better suited for machine learning analytics (e.g., Apache Spark).

    \item \textbf{Vendor lock-in and lock-out:} Typically, an RDBMS-OLAP stores data in its internal proprietary formats, which limits data access and export, and therefore hinders building robust models. One strategy enterprises have been following to enable these workloads is to offload the data (via ETL process) from the RDBMS-OLAP into a cloud data lake, such as Amazon S3 \cite{Mishra11} or Microsoft ADLS. However, this approach adds further complexity in dealing with the ETL pipelines and their failures. It also leads to additional data copies that are often unmanageable and can lead to issues such as data drift, model decay, etc.

    \item \textbf{High costs:} RDBMS-OLAP (on-prem or cloud) have a considerable cost factor. Although cloud data warehousing systems make it easy for users to start, the cost to store large volumes of data and run any computation on that data increases over time\cite{Datanami1}. Also, any pre-aggregated tables and materialized views \cite{10.1145/319757.319787} created within the RDBMS-OLAP result in additional storage costs. From an analytical use case perspective, usually, there is the need to ingest all the data, but since storage costs are high, it limits the analysis scope to only a few datasets. Apart from the monetary costs, a data warehousing approach also brings in other overhead costs, such as creating and managing ETL/ELT pipelines on the part of data engineers, delayed time to insight, and vendor lock-in of data. It is important to note that some cloud RDBMS-OLAP data warehousing systems offer the usage of external open-format tables \cite{Malone11} for storage. However, external table functionality is usually both complicated and non-performant.
\end{enumerate}

\section{Data Lakehouse}
A data lakehouse is a type of data architecture that combines the desirable attributes of an RDBMS-OLAP (data warehousing) and a data lake. It mitigates the challenges seen with both these technologies and brings out the best in both of them. More formally, a data lakehouse architecture is characterized by the following.

\begin{itemize}
    \item \textbf{Transactional support:} Data lakehouses bring in reliability and consistency (ACID properties) in every transaction, such as INSERT or UPDATE, similar to an RDBMS-OLAP. This ensures safe concurrent reads and writes. 

    \item \textbf{Open data:} The basis of a data lakehouse is storing data in open file formats such as Apache Parquet, ORC, etc., and table formats such as Apache Iceberg\cite{Ice11}, Apache Hudi\cite{Hudi1}, and Delta Lake\cite{Delta1}. This enables different analytical workloads to be run by different engines, often from different vendors/providers, on the same data and avoids locking data into a proprietary format.

    \item \textbf{No copy:} A data lakehouse limits data copy as the compute engine can directly access data from the source. The table format layer in a lakehouse architecture adds a logical model and reliable governance on top of the data lake.

    \item \textbf{Data quality and governance:} A lakehouse system focuses on data governance and quality by adopting tried-and-true best practices from the data warehousing world to ensure proper access control and to adhere to regulatory requirements.

    \item \textbf{Schema management:} Data lakehouses guarantees that a specific set schema is respected when writing new data into the data lake. It also facilitates evolving the schema over time without bearing the cost of rewriting the entire table.

    \item \textbf{Scalability:} A data lakehouse is built on the idea of separated storage and compute. It takes advantage of the low-cost storage options of a data lake, which allows storing data of any type (structured, loosely structured, etc.) and volume (order of petabytes) in open file formats such as Apache Parquet. Similarly, any analytical workload (batch, ad hoc SQL, machine learning) can run on the openly stored data and be scaled independently per the requirements.

\end{itemize}

These attributes set the foundational building blocks for data lakehousing and enable all sorts of data warehousing capabilities in a lakehouse. Ultimately, with a lakehouse architecture, the goal is to eliminate the need for a two-tier architecture to run varying analytical workloads and avoid all the complexities and costs. In the next section, we will examine the technical components that form a data lakehouse architecture.

\section{Requirements of Data Lakehousing}
A data lakehouse breaks down an RDBMS-OLAP’s tightly coupled technical components into separate components. At its core, this is enabled by storing data in open file and table formats, which makes data accessible to every compute engine supporting the format. This provides more agility and choice when architecting such a platform.

\subsection{Technical Components of a Data Lakehouse}
A data lakehouse addresses the same technical components that we discussed in the data warehousing section (Section ~\ref{section:tc}), but in a different way. These components, including storage, file and table formats, compute engines, and catalogs form the foundation of any data warehousing or lakehousing architecture. In this section, we will describe how the technical components of a lakehouse meets the requirements defined by an RDBMS-OLAP and the advantages of having a loosely coupled architecture. An overview of the components of a lakehouse is presented in Figure ~\ref{fig:figure_dlh}.

\begin{enumerate}
    \item \textbf{Data storage:} The first component of an open lakehouse architecture is the storage. This is where the data files land after ingestion from various operational systems via ETL processes. Cloud object stores such as Amazon S3, Azure Storage, and Google Cloud Storage support storing any type of data and can scale to effectively infinite volume. These systems are also very cost-effective\cite{Punuru2019}, which is a huge adoption factor compared to data warehousing storage costs.

    \item \textbf{Storage engine:} The storage engine component of a data lakehouse handles data management tasks like compaction, repartitioning, and indexing. These activities optimize data organization in cloud object storages to ensure efficient query performance. Data optimization services like Dremio Arctic\cite{Arctic1} and Tabular\cite{Tabular1} streamline these tasks, while engines such as Apache Spark, Apache Flink and Dremio Sonar\cite{Sonar1} can also be utilized with manual configuration, which can then be automated through orchestration and scheduling tools.

\begin{figure}
  \centering
  \includegraphics[width=\linewidth]{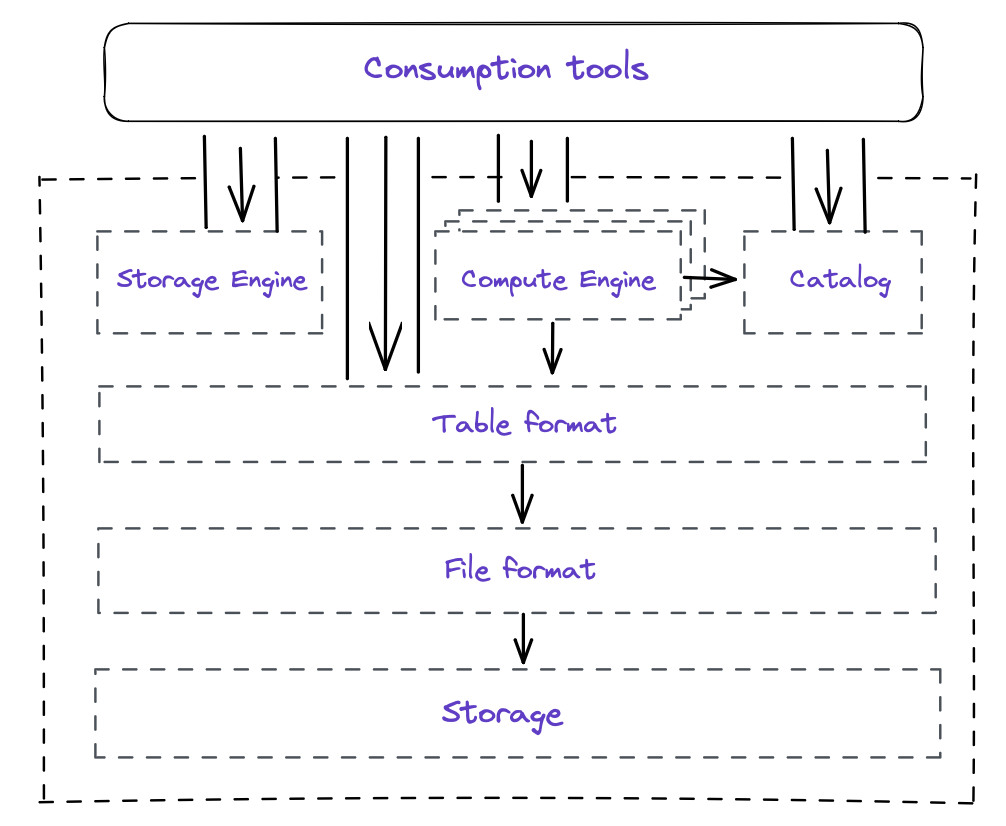}
  \caption{A data lakehouse architecture with the various components}
  \label{fig:figure_dlh}
\end{figure}

    \item \textbf{File format:} File formats hold the actual raw data in a data lakehouse architecture and are stored physically on the object storage. Since the file formats in a data lakehouse are open formats such as Apache Parquet or JSON, they allow data to be consumed by multiple engines for various types of workloads (e.g., one engine optimized for BI, one engine optimized for machine learning). Typically, these are column-oriented, providing significant advantages in reading or sharing data between multiple systems.

    \item \textbf{Table format:} A key component of a data lakehouse architecture is the table format, which acts as a metadata layer atop file formats like Apache Parquet. It simplifies data file organization and management within the data lake, abstracting away the complexity of the physical data structure. By utilizing the APIs provided by the table format, different engines can perform simultaneous reads and writes on the same dataset. This enables the execution of RDBMS-OLAP transactions (DML) with atomicity and consistency guarantees while keeping data in open formats. Table formats such as Apache Iceberg, Apache Hudi, and Delta Lake bring critical capabilities like schema evolution, partitioning, and time travel, previously limited to RDBMS-OLAP systems. We will explore these technical capabilities further in the next section.

    \item \textbf{Catalog:} A catalog, sometimes referred to as a metastore, is a vital component of the lakehouse architecture and enables efficient search and discovery within the data lakehouse. It is a service that keeps track of all the tables in a lakehouse, i.e., the metadata information. The catalog has information about each table's name and schema (column names and types) and a reference to the metadata for each table (table format). Catalogs can be of different types. For example, Apache Iceberg works with catalogs such as HDFS, Hive, AWS Glue, Nessie, Dremio Arctic, DynamoDB, REST, etc.

    \item \textbf{Compute engine:} In a data lakehouse architecture, the compute engine is responsible for processing data and ensuring efficient read and write performance. Table formats provide the necessary specifications and APIs for interacting with the data, while the compute engine handles the actual processing. Compute engines vary depending on the workload type, with options like Dremio Sonar for low-latency ad hoc SQL, Apache Flink for streaming, and Apache Spark for machine learning tasks.
\end{enumerate}

\subsection{Technical Capabilities}
\label{section:tc_lh}
This section reviews the technical capabilities of a system that can support data warehousing established in the “Requirements of Data Warehousing” section (section ~\ref{section:tcp}) and discusses how a data lakehouse addresses each requirement.

\begin{enumerate}
    \item \textbf{Governance and security:} Like RDBMS-OLAPs, data lakehouses store vast amounts of data, emphasizing the need for robust data governance policies to ensure appropriate access. As data resides in raw form in object storage, rather than a predefined structure, securing sensitive data (such as PII, etc.) becomes crucial, necessitating stringent security measures. Lakehouse platforms offer a range of tools and open source solutions to effectively address data governance and security requirements. Apache Ranger\cite{Ranger1}, for instance, is widely used for authorization and access control, supporting role-based filtering and column-level masking. Modern lakehouse catalogs like Dremio Arctic and Tabular provide fine-grained access control and audit logging, ensuring compliance with regulations such as HIPAA and GDPR.

    \item \textbf{High concurrency:} Similar to RDBMS-OLAPs, lakehouse platforms allow scaling workloads and handling concurrency levels while controlling the compute resources (limit costs, etc.). For example, lakehouse platforms like Dremio Cloud and Databricks brings in capabilities for both planning and execution resources to dynamically scale based on workload demand\cite{Dremio22}, thereby allowing to deal with any level of concurrency while maintaining consistent performance. The implementation of table formats such as Apache Iceberg in a lakehouse architecture facilitates concurrent reads and writes using mechanisms such as optimistic concurrency control. Moreover, by enforcing ACID compliance, table formats ensure that downstream applications are not affected by potentially "bad" data resulting from concurrent writes.

    \item \textbf{Low query latency:} As in an RDBMS-OLAP, compute engines in a data lakehouse architecture apply various optimization techniques such as partitioning, compaction, clustering, indexing, etc., facilitated by table formats together with the engines’ query acceleration technologies to support low-latency queries. Building and scaling BI dashboards based on cloud data lakes have always been challenging. However, modern lakehouse platforms bring in capabilities to create high-performance dashboards directly against the data in the lake using query acceleration technologies. For example, Dremio Sonar provides data reflections, which are transparently substituted materialized views on heterogeneous source systems. Similarly, Databricks SQL provides support for creation of materialized views that store pre-computed results based on the latest data from source table to improve query performance. 

    \item \textbf{Ad hoc queries:} A data lakehouse platform includes compute engines that support ad hoc queries, allowing users to directly query data from the data lake storage. Also, as opposed to an RDBMS-OLAP, data consumers such as decision scientists have direct access to the data in the lake, so they have access to more than just a specific narrow curated dataset, which helps them to make their analysis robust. Some data lakehouse compute engines provide native connectivity with BI tools such as Tableau and Power BI, which are used extensively for ad hoc questions and exploratory analytics, for ease of use. Additionally, they also have support for raw SQL.

    \item \textbf{Workload management (WLM):} Like RDBMS-OLAP, data lakehouse compute engines handle a range of workloads, including ad hoc queries and scheduled reporting. In a lakehouse architecture, different engines cater to specific workload types (e.g., interactive dashboarding versus machine learning), ensuring physical isolation. Within each compute engine, workload isolation is achieved through queue configuration, specifying attributes like memory limits, CPU priority, and queueing and runtime timeouts. Furthermore, rules are established to assign queries to their respective queues, optimizing workload management and resource allocation. Workload management facilitates isolated workloads for different users and groups, providing controls for efficient utilization of cluster resources.

    \item \textbf{Schema and physical layout evolution:} The ability to incorporate additional dimension fields or modify the internal structure of existing tables is crucial in a lakehouse architecture, mirroring the capabilities found in RDBMS-OLAP systems. It is essential to support these changes seamlessly, without excessive overhead or the need to rewrite entire production tables. The table formats used in a data lakehouse architecture specifically address this requirement by providing effective schema evolution capabilities. Formats like Apache Iceberg, Hudi and Delta Lake enable in-place table evolution using SQL operations such as Add, Drop, and Rename, which only affect metadata and do not require data file rewriting or table migration. Additionally, Apache Iceberg tables allow for the evolution of partition layouts to accommodate changes in data volumes. With features like hidden partitioning\cite{Hidden1}, partitions can be modified without significant impact on previously written queries.

    \item \textbf{ACID-compliant transactions:} As you would expect in an RDBMS-OLAP, one of the essential attributes of a lakehouse is the support for ACID-compliant queries. Until now, with data lakes, there was not a practical way at scale to ensure a transaction was reliable when multiple users wrote data simultaneously\cite{Arch1}. This impacts the downstream applications that depend on the data. The ACID guarantees facilitated by the table formats in a lakehouse bring consistency and correctness in every transaction, allowing several engines to work concurrently. Therefore, operations such as incremental loads and modifying data (updates, deletes) can succeed concurrently in a lakehouse. ACID-based transactions are pivotal for reliable data across data products (dashboards, ML models, etc.).
    
    Here are some of the key features related to transactions in a lakehouse.

    \begin{itemize}
        \item Multi-statement \& Multi-table transactions: Data lakehouse compute engines enable multi-statement transactions, allowing for mutating operations such as inserting or deleting rows across multiple tables. Table formats like Apache Iceberg provide ACID-compliant support for multi-statement transactions, ensuring atomicity and consistency. Multi-statement transactions across multiple tables are enabled by solutions such as Project Nessie\cite{Nessie1} and LakeFS\cite{LakeFS} that act as version control for table formats and object stores, making it possible to isolate changes and merge changes atomically.

        \item Transaction rollback: The ability to roll back transactions and go back to a previous version of a table is another important capability of an RDBMS-OLAP system. In a lakehouse, table formats such as Apache Iceberg maintain a historical lineage of tables state via snapshots. Snapshots allow querying prior table states and help achieve transaction rollbacks to quickly correct any data in situations that demand it.

        \item Locking model: Similar to data warehousing systems, there can be scenarios when multiple users or engines would like to write to the same table simultaneously in a lakehouse. Data lakehouse table formats ensure concurrent writes are safe and do not result in bad data across applications. Concurrency models, such as optimistic concurrency control\cite{OCC}, makes it possible to achieve transactional guarantees in all the three table formats, i.e. Apache Iceberg, Hudi and Delta Lake.
    \end{itemize}

    \item \textbf{Separation of storage and compute:} While the separation of storage and compute has been a relatively recent development for RDBMS-OLAP systems, this fundamental principle has long been ingrained in the data lakehouse architecture. As a vast repository to run diverse workloads like ad hoc SQL, ETL, and machine learning, a data lakehouse inherently has the ability to independently scale its storage and compute components according to specific requirements, owing to the massive data volumes it encompasses.
\end{enumerate}

\subsection{Technology-Independent Practices}
The technology-independent practices listed below highlight how a data lakehouse architecture supports the critical practices usually employed in the data warehousing world.

\begin{enumerate}
    \item \textbf {Data modeling:} Just like in data warehousing, data modeling is vital in the lakehouse architecture. It defines the model of real-world data, capturing relationships between data objects. Data modeling is technology-agnostic, i.e., the methods to model data and bring it to a standard form to answer any business problems fairly remain the same across architectures. The lakehouse platform supports various modeling techniques such as star and snowflake schemas, separating physical data from logical representation. It follows common semantic layer approaches, organizing data into layers like physical (raw), staging (cleansed), business (transformed per requirement), and application (accessible to apps). Data integration consolidates data from diverse sources, each with its unique structure, into a cohesive format in the physical layer, while the business layer enables further transformations based on modeling methods. This unified data empowers users to extract insights and make informed decisions.

    \item \textbf{ETL/ELT:} In the traditional data warehousing world, a common approach is extracting data from various sources, transforming it, and then loading it into the RDBMS-OLAP storage. In contrast, with a lakehouse architecture, the idea is first to extract and load the data to a data lake and then perform the necessary transformations using an engine of your choice. This provides the advantage of having “schema-on-read,” \cite{3bf9e596e1264cdfb4d455f17cd705d9} which means you don’t have to worry about processing at this stage and be able to land data in raw formats. Once the data is landed, transformation logic can be applied to provide “schema-on-write” benefits. Then, depending on the analytical workload (ad hoc SQL, ML, etc.), data can be processed accordingly (by engines such as Spark and TensorFlow) and made available for consumption by the various downstream applications (BI tools, notebooks). Recently, the industry has been leveraging this ELT approach more with RDBMS-OLAP systems, however, this approach doesn’t enable landing any data in a “schema-on-read” approach as well, and there are usually high costs associated with this approach on RDBMS-OLAP systems.

    \item \textbf{Data quality:} Maintaining high data quality is a fundamental requirement in a data lakehouse architecture. With the diverse nature of data in a lakehouse, ensuring accuracy, consistency, and reliability becomes paramount. Three key practices contribute to maintaining data integrity and accuracy.

    \begin{itemize}
        \item Master Data Management (MDM): Like any other data platform, MDM practices can be applied to the data in a lakehouse to ensure the accuracy and consistency of core data assets (master data). This is essential for critical data assets like BI reports, where accurate information is crucial for decision making. While not all data types may be suitable for creating master data entities, evaluating the applicability of MDM practices based on the analytical use cases can provide expected benefits.

        \item Referential integrity: Similar to traditional data warehousing, data lakehouse architectures leverage compute engines and table formats to enforce referential integrity checks, thereby ensuring the accuracy and consistency of data relationships and schema. For example, SQL-based compute engines allow performing explicit join queries between various fact and dimension tables to get the lookups of the fields and foreign key relationships within the fact table. This way any missing keys can be detected to avoid wrong results in downstream applications. Additionally, branching capabilities facilitated by lakehouse version control technologies like Project Nessie and LakeFS makes it possible to do these checks in isolation and only expose changes to users after this validation has been done.

        \item Slowly changing dimensions: Implementing slowly changing dimensions is a common practice in data warehousing to preserve historical changes and ensure no downstream applications are impacted due to the changes. Table formats like Apache Iceberg, Hudi and Delta Lake brings these capabilities to a data lakehouse. Features such as ACID-compliant queries and row-level updates/deletes, enable the implementation of SCDs (e.g., SCD type 2, SCD type 4).

    \end{itemize}
\end{enumerate}

\section{Additional Value That Data Lakehousing Brings}
Up to this point, we have discussed the various requirements of a data lakehouse architecture that help to achieve similar capabilities as a data warehousing system. In this section we list the additional values that a lakehouse architecture offers.

\begin{enumerate}
    \item \textbf{Open data architecture:} One of the most significant values of adopting a data lakehouse platform is enabling an open data architecture. With such an architecture, data is stored as an independent tier in open file formats (e.g., Parquet) and table formats (e.g., Apache Iceberg) in an organization’s own cloud storage. This allows multiple analytical engines to access data directly and run together in unison based on the different types of workloads. This is important because you can leverage different compute engines today for the workloads they were designed to address, but also leave yourself open to adopting new compute engines tomorrow for future workload needs. These future compute engines may not even exist yet. An open architecture eliminates the need to lock your data to a vendor-specific format (unlike RDBMS-OLAP) to be leveraged by one specific compute engine and opens up possibilities for all sorts of analytical use cases directly on the open data.

    \item \textbf{Fewer data copies and better governance:} In a traditional two-tier data architecture, data is first moved from various operational systems to a data lake using a combination of ETL, change data capture (CDC) \cite{8275102}, and streaming tools. This data is stored in its raw format within the “landing zones.” Subsequently, the data undergoes a series of processing steps to transform it into a structured format that is suitable for data consumers. For interactive BI workloads, for example, performance is critical. In this traditional two-tier architecture, the data needs to be moved to an RDBMS-OLAP, thereby creating more data copies. Furthermore, RDBMS-OLAPs often don’t offer sufficient performance when dealing with large amounts of granular data, necessitating copying and transforming it into aggregation tables, resulting in additional time and resources spent, as well as additional data copies. Whereas in a lakehouse architecture, interactive data lake compute engines powered by query acceleration technologies allows data to be queried directly on the lake, thereby eliminating the need to copy the data into the RDBMS-OLAP. Further, the ability to create and incrementally refresh materialized views in certain lakehouse platforms help minimize the cost of data copies needed for performance.
For machine learning-based tasks, you might require specific copies of data present in the RDBMS-OLAP to be exported to external file formats (such as Parquet), which the ML frameworks can then consume. Making data copies results in complexities and challenges, such as managing many ETL/ELT pipelines, unmanageable data copies, and issues like data drift, etc. 
Additionally, since the data teams and stewards have little to no control over these extra data copies, there is no governance around the data, which can pose a big risk. With a data lakehouse architecture, data is stored in open file and table formats accessible to various analytical engines. This allows running workloads such as reporting and machine learning directly on the stored data in a cloud data lake, minimizing the data copies, and therefore maximizing proper governance. 

\item \textbf {Manage data as code:} The data lakehouse brings new capabilities to manage data using a code-based approach. Enabling this paradigm shift are catalogs like Project Nessie and data lake version control systems like LakeFS, which, combined with table formats such as Apache Iceberg or Delta Lake, enable versioning your tables, which empowers data teams with comprehensive control and visibility over their data assets. Such an approach is currently not possible in an RDBMS-OLAP system because of the tightly coupled architectural design and because the table formats and related metadata are accessible only to the specific warehouse engine. Data as code systems serve as a centralized repository that tracks and manages the history of changes made to the tables and its metadata. Similar to version control systems like Git\cite{Git1}, a commit in these systems capture the state of the world at a particular point in time. Each commit represents a specific change or set of changes to the metadata. This new approach makes it possible to isolate data environments (such as creating isolated branches) and conduct experiments, reproduce specific scenarios, perform data quality checks, etc., in a much more flexible and efficient way. For example, a distributed batch job writing data into a table can do so in isolation, and no downstream application has to see any incomplete data. These changes can be exposed to a small set of users and/or applications, gradually increasing the rollout until there is confidence that the rollout is production-ready. Then, the new changes can be atomically merged and become visible to all users. This is similar to the effective blue-green deployment model\cite{Blue11} employed in the deployment of software. 

\item \textbf {Federation:} A data lakehouse architecture enables federation, allowing data consumers to access and utilize data beyond a single storage system. Unlike traditional architectures, where data retrieval from different systems involves time-consuming ETL processes, a lakehouse platform (with its compute engine) provides direct access to diverse datasets. For example, a data scientist working on a churn prediction model using data from a data lake can seamlessly incorporate high-quality business datasets stored in a system like RDBMS-OLAP without the need for data transfers or delays. This streamlined approach enhances collaboration, accelerates analysis, and empowers users to leverage data from multiple sources for informed decision making.
\end{enumerate}

\section{Example of a Lakehouse Implementation}
The modular and open nature of a data lakehouse architecture gives us the flexibility to choose the best-of-breed engines and tools depending on the requirements. So, the implementation of a lakehouse can vary from use case to use case. This section presents a basic framework of a lakehouse architecture with Apache Iceberg as the table format catering to two common analytical workloads: BI and machine learning. The choice of Apache Iceberg as the table format is based on its wide community support and to benefit from key features such as hidden partitioning and schema evolution\cite{tagliabue2023building}. We acknowledge that the complexity (workloads, security, etc.) and tool stack will vary for large-scale implementations. The idea of this section is to serve as a guide for architects and engineers to get started with lakehouse systems. Similar implementations can also be achieved with Apache Hudi or Delta Lake as the table format and engines such as Trino or Apache Flink.

\begin{figure} [h]
  \centering
  \includegraphics[width=1\linewidth]{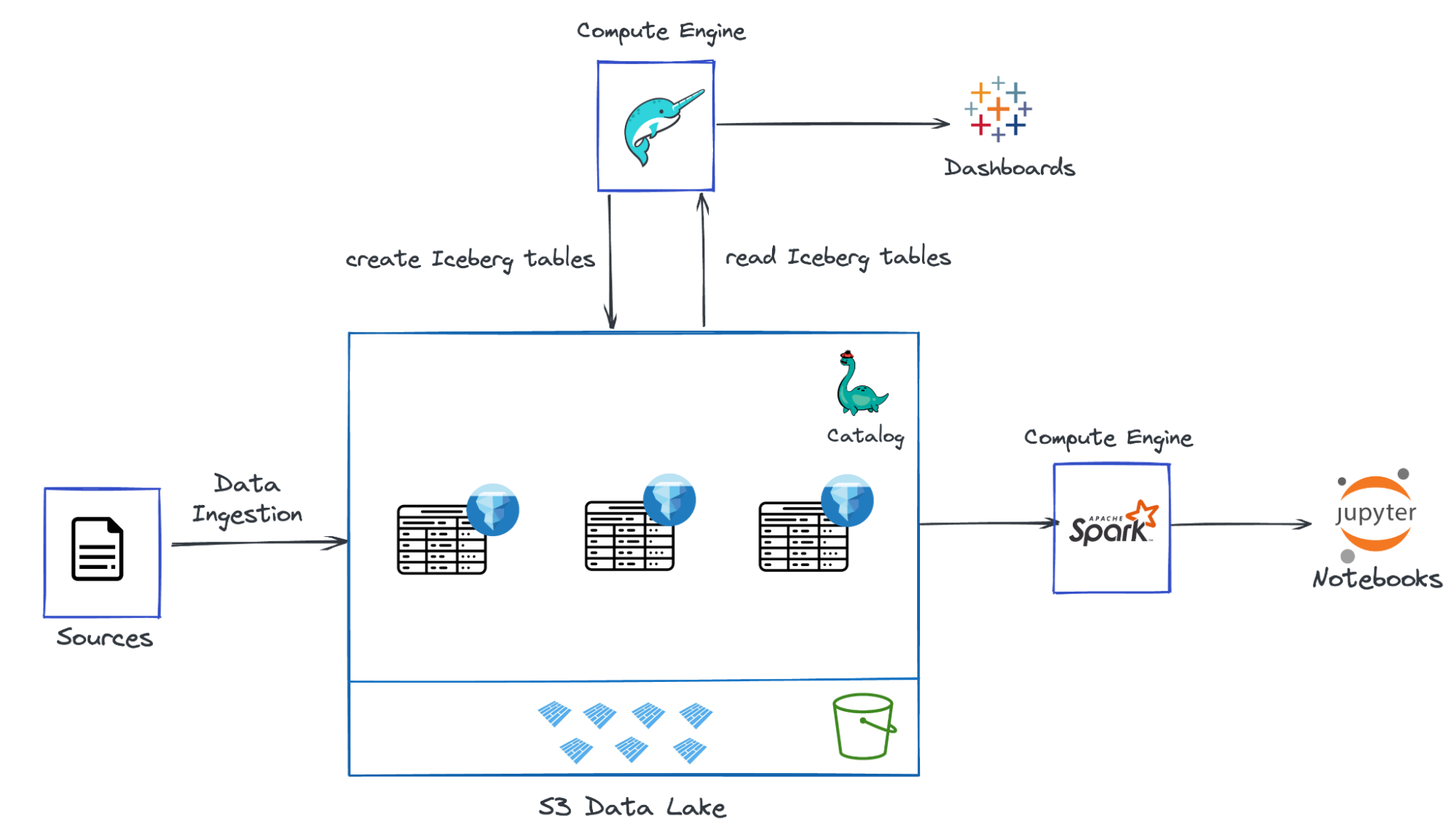}
  \caption{An example data lakehouse implementation}
  \label{fig:figure_arch}
\end{figure}

Our first goal is to have a lakehouse system to run analytical reports on top of it. An example architecture is presented in Figure ~\ref{fig:figure_arch}. 

We will start with ingesting all of our data in a cloud object store, which in this case is Amazon S3. The data files we ingest into our data lake are all in Apache Parquet format. Now, to have the ability to run ACID transactions on our data files and benefit from all the data management capabilities as in an RDBMS-OLAP, we will use Apache Iceberg as our table format. Iceberg serves as a metadata layer and helps us have that abstraction on our physical data files. Apache Iceberg also brings in various file optimization strategies and maintenance methods out of the box, so we can ensure our queries remain fast and performant. The next step is to add a catalog that can track all the tables in our lakehouse and hold the metadata information (table name, schema, etc.). There are various options in terms of the catalog we want to use with Apache Iceberg. For this specific use case, we will use the Project Nessie catalog, which is an open source solution. Nessie also brings in Git-like abilities to data lakes so we can have isolation for different data environments (dev, prod), which can be extremely beneficial as we mature with our lakehouse architecture and add more workloads in the future. The next important component in our architecture is the compute engine which will help us process the data files managed by the Iceberg table and let us perform transactions (based on our workload). Dremio Sonar, a SQL-based distributed query engine, will be used for this. \\

We connect to our data source (S3 bucket) from Dremio Sonar and create an Iceberg table using SQL.


\begin{lstlisting}[label={lst:code}, breaklines=true, postbreak=\mbox{\textcolor{red}{$\hookrightarrow$}\space}]
CREATE TABLE churnquarter AS
SELECT * FROM "churn-bigml-20_allfeat_Oct_train_data.parquet"
\end{lstlisting}

\begin{figure} [h]
  \centering
  \includegraphics[width=\linewidth]{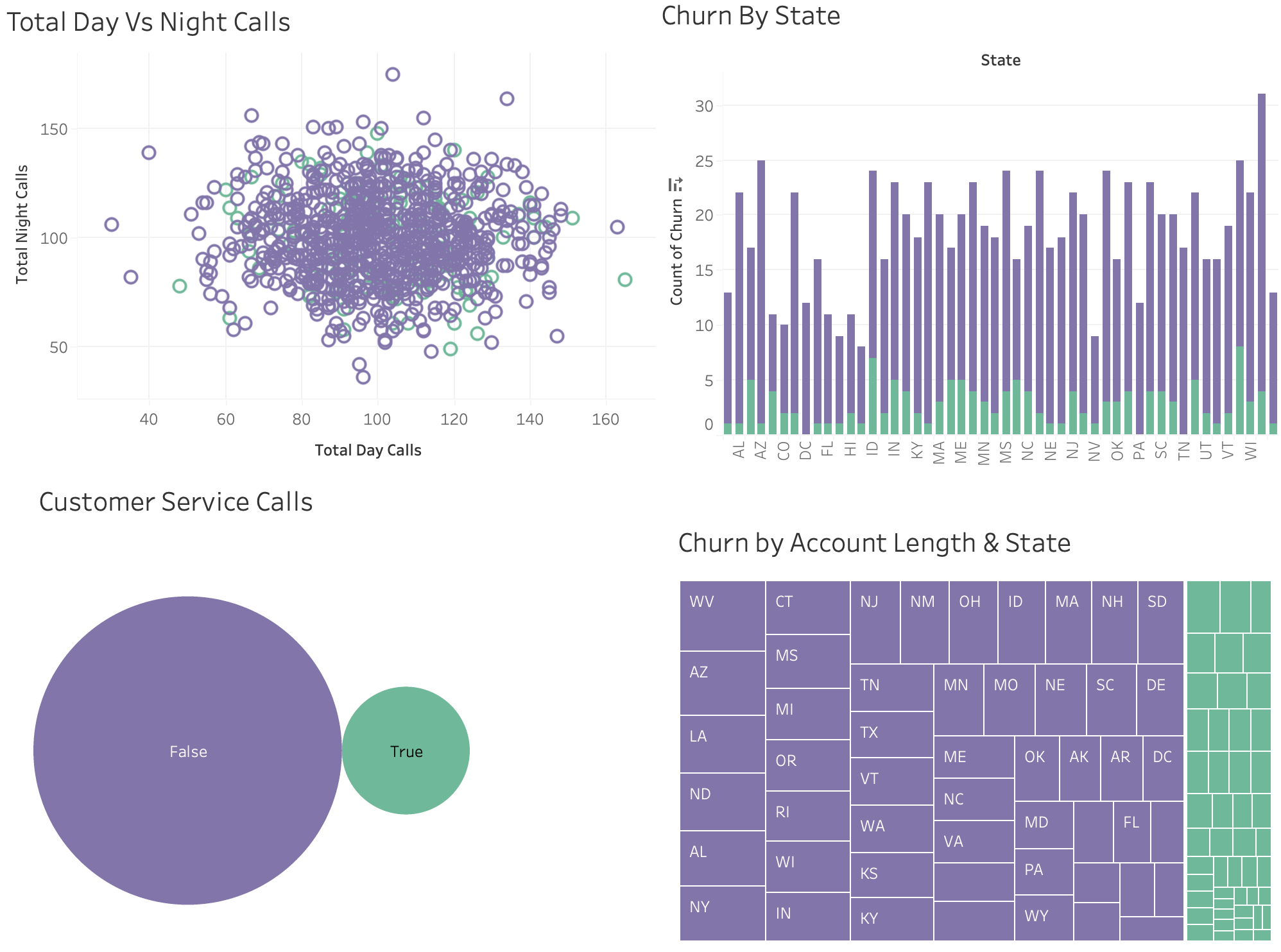}
  \caption{A dashboard built on top of an Apache Iceberg table.}
  \label{fig:figure_dash}
\end{figure}

As discussed in Section ~\ref{section:tc_lh}, some lakehouse platforms enable native connection to tools such as Tableau. This provides a faster and easier way to run analytical workloads such as BI directly on the data lake. Figure ~\ref{fig:figure_dash}  presents a dashboard that leverages Dremio’s compute engine to run queries using live query mode on an Apache Iceberg table \textit{churnquarter}.

For the next use case, we will train a churn classification model using the same Iceberg dataset \textit{churnquarter} that we created in the previous use case. Since this dataset is directly available on the data lake, we eliminate the need for exporting and creating or maintaining ETL pipelines. And since this data is stored in open table format like Apache Iceberg, it is directly accessible by other compute engines. The compute engine and ML framework that we use here is Apache Spark and scikit-learn\cite{Sk1}.


\begin{lstlisting}[label={lst:code}, breaklines=true, postbreak=\mbox{\textcolor{red}{$\hookrightarrow$}\space}]
df_telco = spark.read.table("arctic.telco.churnquarter")
target = df_telco.iloc[:, -1].values
features = df_telco.iloc[:, :-1].values
from sklearn.model_selection import train_test_split
X_train, X_test, y_train, y_test = train_test_split(features, target, test_size=0.20, random_state=101)
rfc = RandomForestClassifier(n_estimators=600)
rfc.fit(X_train, y_train)
predictions = rfc.predict(X_test)
acc = accuracy_score(y_test, predictions)
classReport = classification_report(y_test, predictions)
confMatrix = confusion_matrix(y_test, predictions)
print('Evaluation of the trained model:')
print('Accuracy:', acc)
print('Confusion Matrix:\n', confMatrix)
print('Classification Report:\n', classReport)
\end{lstlisting}

\begin{figure} [h]
  \centering
  \includegraphics[width=\linewidth]{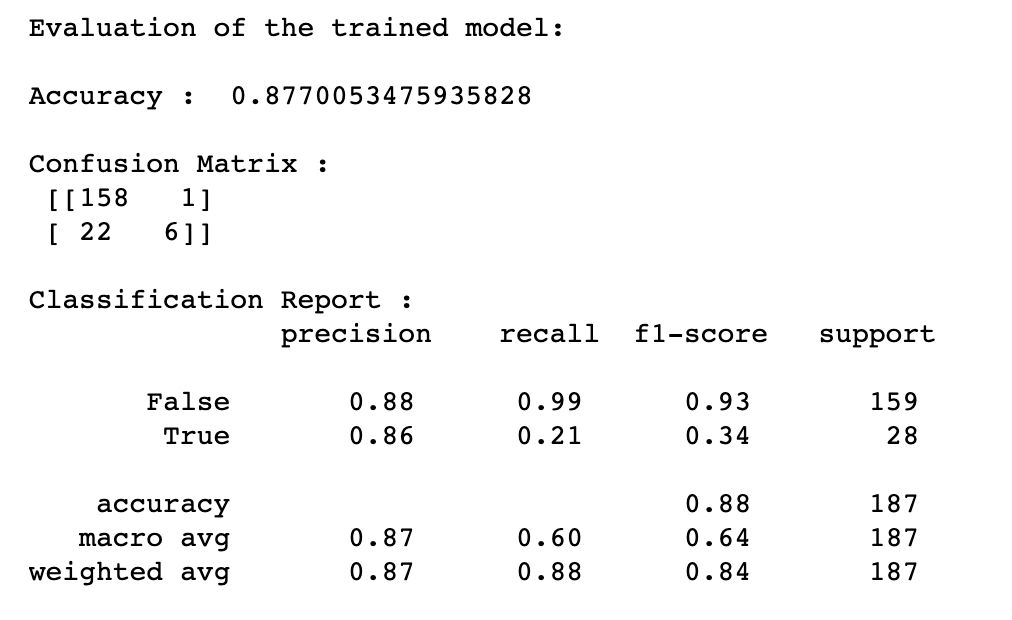}
  \caption{Results from the classification model }
  \label{fig:figure_label}
\end{figure}

Figure ~\ref{fig:figure_label} shows the evaluation for the churn classification model.

In this section, we have showcased the inherent versatility of a data lakehouse architecture, which allows for concurrent execution of diverse analytical workloads, including BI and machine learning, on a shared dataset. This eliminates the necessity of data movement, costly duplication, and intricate management of ETL pipelines. Consequently, this unified architecture offers a streamlined approach to address the analytical objectives of any organization.

\section{Conclusion}
In this paper, we presented the various components that make up a traditional data warehousing system with the technological component (RDBMS-OLAP) and the technology-independent practices associated with such a platform. We then argued and described how a data lakehouse provides equivalent technical and data management capabilities and supports practices such as data modeling, data quality, ETL/ELT, etc., with a much more robust architecture. The open nature of a lakehouse architecture eliminates locking data into a vendor-specific proprietary format and makes it possible for multiple analytical engines to coexist in the same ecosystem. This also makes the data architecture future-proof, allowing you to adopt new tools based on the requirements. The scalability and agility of cloud object stores in a lakehouse architecture enables businesses to store huge volumes of data with varying types (structured, semi-structured, etc.) without having to worry about significantly high costs, unlike cloud data warehousing systems. In summary, the paper has explored the capabilities of a data lakehouse architecture and demonstrated that it encompasses the core functionalities of a traditional data warehousing system while offering additional benefits.

\bibliographystyle{ACM-Reference-Format}
\bibliography{sample-base}

\end{document}